\documentclass[a4paper]{article}

\usepackage{INTERSPEECH2020}
\usepackage[ruled,vlined]{algorithm2e}
\usepackage{multirow}
\usepackage{xcolor,soul}
\usepackage{adjustbox}
\usepackage{url}

\title{Bunched LPCNet : Vocoder for Low-cost Neural Text-To-Speech Systems}

\name{Ravichander Vipperla$^{1}$\renewcommand{\thefootnote}{\fnsymbol{footnote}}\footnote[2]{Indicates equal contribution.}, Sangjun Park$^{2}$\footnote[2]{}, Kihyun Choo$^{2}$\footnote[2]{}, Samin Ishtiaq$^1$, Kyoungbo Min$^2$\\Sourav Bhattacharya$^1$, Abhinav Mehrotra$^1$, Alberto Gil C. P. Ramos$^1$, Nicholas D. Lane$^{1,3}$ }
\address{
  $^1$Samsung AI Centre, Cambridge, UK \\
  $^2$Samsung Research, Seoul, Republic of Korea \\ 
  $^3$University of Cambridge, UK}
\email{\{r.vipperla, sj0.park, khchoo, s.ishtiaq, kyoungbo.min,\\ sourav.b1, a.mehrotra1, a.gilramos, nic.lane\}@samsung.com}

\graphicspath{ {./figures/} }

\begin{document}

\maketitle

\begin{abstract}

LPCNet is an efficient vocoder that combines linear prediction and deep neural network modules to keep the computational complexity low. 
In this work, we present two techniques to further reduce it's complexity, aiming for a low-cost LPCNet vocoder-based neural Text-to-Speech (TTS) System. 
%
These techniques are: 
1) Sample-bunching, which allows LPCNet to generate more than one audio sample per inference; and 
2) Bit-bunching, which reduces the computations in the final layer of LPCNet. 
%
With the proposed bunching techniques, LPCNet, in conjunction with a Deep Convolutional TTS (DCTTS) acoustic model, shows a 2.19x improvement over the baseline run-time when running on a mobile device, with a less than 0.1 decrease in TTS mean opinion score (MOS).


\end{abstract}

\renewcommand{\thefootnote}{\fnsymbol{footnote}}\footnotetext{\footnote[2]{}Indicates equal contribution.}
\renewcommand{\thefootnote}{\arabic{footnote}}


\section{Introduction}
LPCNet~\cite{valin-lpcnet} is a state-of-the-art light-weight vocoder that improves upon WaveRNN~\cite{kalchbrenner2018efficient} in terms of sound quality and inference time. Its design is based on the principles of source-filter model~\cite{sound-principles} of speech production. The key idea in LPCNet is to separate the burden of vocal tract response prediction using a well-understood, low-cost, linear prediction filter and utilize a smaller, WaveRNN-style neural net's capacity for the prediction of source excitation in order to reconstruct the speech.  It is one of the most compute efficient neural vocoders in recent times and has been demonstrated to give excellent performance when used in text-to-speech systems~\cite{kons2019high, hwang2020improving} as well as in speech codecs~\cite{valin-lpcnet-interspeech19}. 
Due to its small size and low computational complexity, it is well suited as the vocoder component for commercial on-device TTS solutions for mobile and IoT devices.
An auto-regressive vocoder in a TTS system is the biggest computational bottleneck as it infers speech samples one at a time conditioned on the previously generated samples. For a wide-band speech signal generation at 16 KHz or 24 KHz, even an efficient LPCNet implementation accounts for 80 to 90\% of the total computational cost, the rest accounted for by the NLP and acoustic modeling components that generate the input for the vocoder.



In this paper we address the problem of reducing the computational complexity of LPCNet without sacrificing the synthesized audio quality. Our contributions include two methods: 1) Sample bunching where the architecture of the LPCNet has been modified to generate more than one sample per inference and 2) Bit bunching where the final, softmax layer is segregated into two bunches to reduce the layer size and computations.
 
We present a brief overview of LPCNet in Section 2, followed by in-depth description of the sample and bit bunching techniques in Sections 3 and 4 respectively. Evaluation results highlighting the efficacy of the proposed methods are presented in Section 5.

\section{LPCNet overview}
\label{section:lpcnet}
\begin{figure}[h]
\centering
\includegraphics[width=0.4\textwidth]{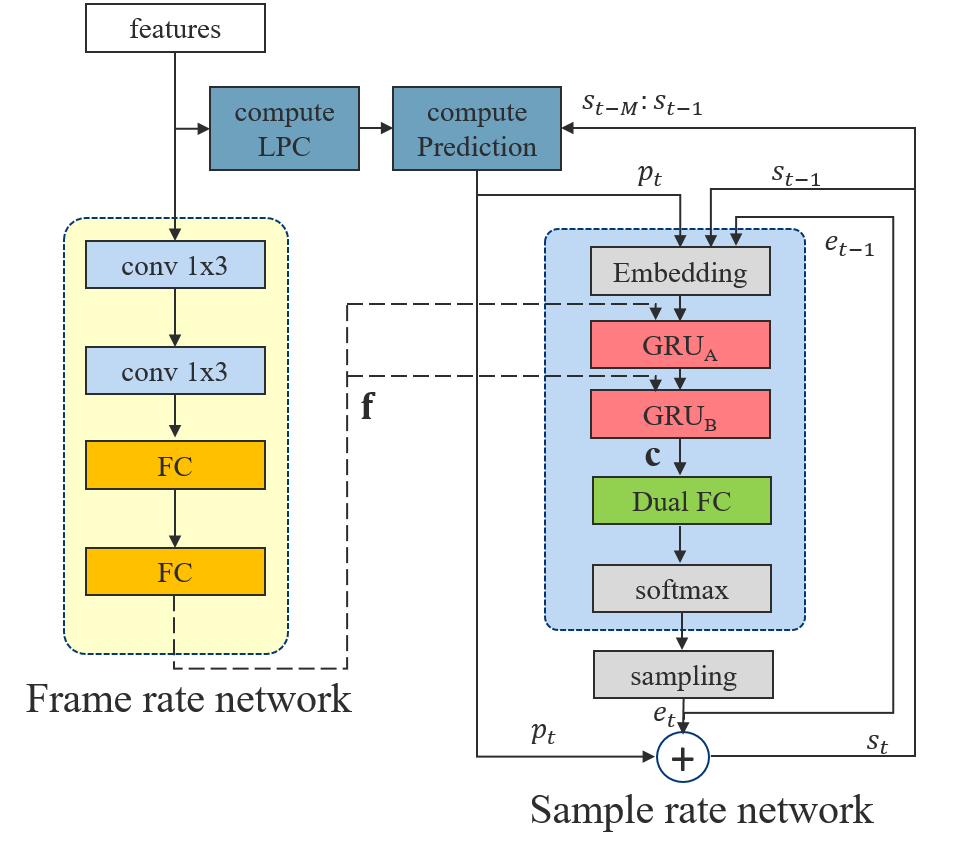}
\caption{LPCNet overview}
\label{fig:lpcnet_overview}
\end{figure}

LPCNet, as depicted in Figure~\ref{fig:lpcnet_overview}, keeps the computational complexity low by using an all-pole LPC filter ($M$ co-efficients)~\cite{makhoul-lpc} for modeling the vocal tract response and a small neural network for predicting the excitation signals. It comprises a frame rate network (FRN) that runs once per input frame, and a sample rate network (SRN) that runs $N$ (frame size)  times per frame generating one audio sample per inference. As a result, most of the computational burden resides in the SRN.

The SRN comprises two gated recurrent units (GRU) and one dual fully-connected (dual FC) layer leading to a softmax layer that models the probability distribution of the excitation signal. The excitation signal $e_t$ is sampled from this distribution and combined with the prediction $p_t$ from the LPC filter to generate the audio sample $s_t$. The excitation and speech sample from the previous time step along with the prediction for the current time step are fed as input to the SRN via embedded representations.

Weight sparsification~\cite{kalchbrenner2018efficient} applied to the recurrent weight matrix in $\text{GRU}_{\text{A}}$ can reduce the complexity without quality degradation. Notably, in $\text{GRU}_{\text{A}}$, while the complexity of the input weight matrix $U$ is larger than that of the recurrent weight matrix $W$, matrix-vector multiplications of $U$ with the embedding vectors from $p$, $s$ and $e$ are converted into addition operations using pre-computed lookup tables obtained from multiplication of the embedding tables and the corresponding weights. With these optimizations, LPCNet shows high performance with much lower complexity as compared to other auto-regressive (AR) neural vocoders~\cite{kalchbrenner2018efficient, oord2016wavenet}.

We note that within the SRN, the two GRU units and the dual FC layer account for about 85\% and 15\% of the computations respectively. The two proposed techniques, sampling bunching and bit bunching, target reducing computations in these blocks respectively.


\section{Sample bunching}
\label{section:samplebunching}
The key idea with sample bunching is to get the SRN to generate more than one sample (called a bunch herewith) per inference thereby allowing it to run fewer times resulting in a reduction of computational cost.


The concept of generating multiple samples has been tried in the context of non-autoregressive vocoders~\cite{pmlr-v80-oord18a,prenger2019waveglow, peng2019parallel, yamamoto2020parallel,clarinet2018} via parallel inference on GPUs. Such models do not scale well for CPU processing. In AR vocoders, parallel sample generation is challenging due to the dependence of current inference on past output. SampleRNN~\cite{mehri2018samplernn}, an efficient AR model, uses different clock rates for higher and lower layers in the network to increase throughput. In recently proposed Gaussian LPCNet~\cite{9053337}, two samples are generated per inference by assuming independence for these samples. In order to maintain the quality of output, the capacity of RNN was increased in that work. 

In our approach described below, we maintain the auto-regressive nature of LPCNet in conditioning the current output on past outputs and thereby enabling the maintenance of audio quality beyond bunch size of 2. Our approach allows multi-sample generation on any hardware including low-end CPUs.







\begin{figure}[h]
\centering
\includegraphics[width=0.45\textwidth]{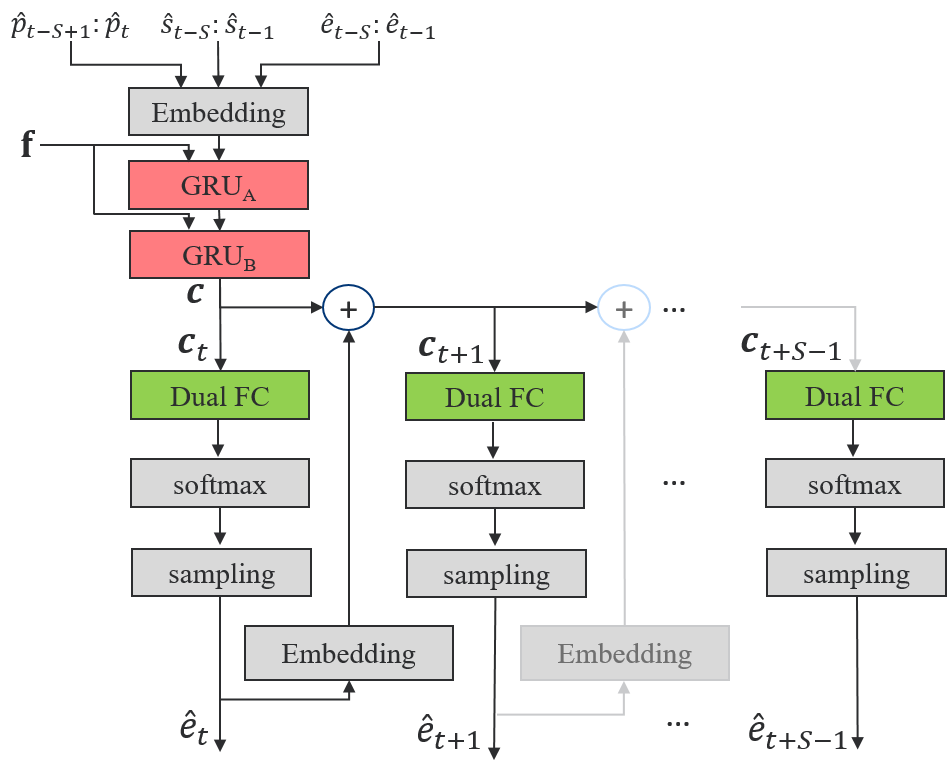}
\caption{Sample bunching}
\label{fig:sample_bunching}
\end{figure}

Our proposition is that the GRUs in the SRN have sufficient model capacity to generate a bunch of samples. As seen in Figure~\ref{fig:sample_bunching}, the SRN shares the GRU layers for all the samples in the bunch and has an individual dual FC layer for each excitation prediction in the bunch. The input to the dual FC layer for the first excitation is conditioned only on the output from $\text{GRU}_{\text{B}}$, \(\hat{e}_{t} \sim p(e_{t}|\bm{c})\); while for the rest, it is also conditioned on the previous excitations within the bunch via embedding feeds, \(\hat{e}_{t+k} \sim p(e_{t+k}|\bm{c}, \hat{e}_{t}, ..., \hat{e}_{t+k-1})\).   



While the number of iterations for the SRN is reduced by the bunch size $S$, the inputs to $\text{GRU}_{\text{A}}$ increase linearly with $S$ resulting in larger input matrix $U$ in $\text{GRU}_{\text{A}}$. However, using the lookup table implementation as described in Section \ref{section:lpcnet}, this increase in cost is marginal and the overall gain in computations with sample bunching is proportional to $1/S$.

\section{Bit bunching}
\label{section:bitbunching}

LPCNet~\cite{valin-lpcnet} uses a Dual FC layer with a softmax activation layer of size 256 for computing the probability $p(e_{t})$. Each softmax output node corresponds to a quantized level of the signal in 8-bits $\mu$-law representation. Inspired by the dual softmax layer in WaveRNN~\cite{kalchbrenner2018efficient}, we introduce the idea of bit bunching for this layer to optimize the inference speed further in conjunction with sample bunching. 

Modifying the Dual FC layer to split 8 bits into two separate groups, higher bit and lower bit bunches, results in two smaller output layers and thereby gains in computational complexity. Our bit bunching approach also uses a new scaled $\mu$-law quantization to allow higher number of bits for better quality signal.

\begin{figure}[h]
\centering
\includegraphics[width=0.25\textwidth]{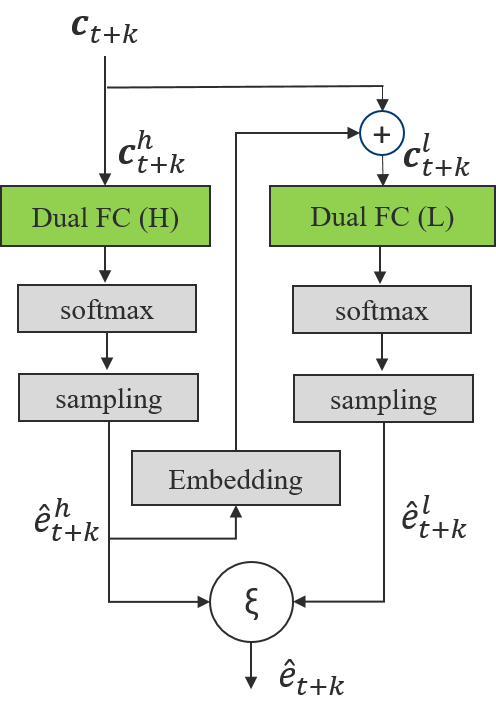}
\caption{Bit bunching}
\label{fig:bit_bunching}
\end{figure}

The idea of bit bunching is depicted in Figure~\ref{fig:bit_bunching}. With bit bunching, the input information into the Dual FC layer is not changed from the original LPCNet, i.e., the information for predicting probability $p(e_{t})$ remains the same. The higher bit bunch and lower bit bunch outputs map to a coarse prediction and fine correction of the excitation signal respectively. To improve the prediction efficacy of the lower bits, the higher bit predictions are additionally fed in as conditioning input via embedding layer. In our experiments, this additional conditioning was found to be useful in improving the cross-entropy loss for the lower-bit bunch and also assisted in the choice of number of bits to assign for higher-bit bunch. The best cross entropy error was achieved with a split of $\bm{B}=(B_{h},B_{l})=(7, 4)$ where $B_{h}$ and $B_{l}$ denote the number of bits in higher bit and lower bit bunches respectively, and this split has been used in all further experimentation.
From the higher and lower bit bunch predictions $\hat{e}_{t}^{h}$ and $\hat{e}_{t}^{l}$, the excitation signal is calculated as $\hat{e}_{t} = \xi(\hat{e}_{t}^{h},\hat{e}_{t}^{l}) = 2^{B_{l}}\hat{e}_{t}^{h}+\hat{e}_{t}^{l}$.
From the generated excitations, the predictions and audio samples are computed with sample and bit bunching according to Algorithm \ref{alg:bunch_sample_generation}.

\begin{algorithm}[t]
\LinesNumbered
\SetAlgoLined
\KwIn{Samples, excitations and predictions from previous run of SRN $\hat{s}_{t-S}\colon\hat{s}_{t-1}$ , $\hat{e}_{t-S}\colon\hat{e}_{t-1}$ and $\hat{p}_{t-S+1}\colon\hat{p}_{t}$ respectively }
\KwOut{$\hat{s}_{t}\colon\hat{s}_{t+S-1}$ , $\hat{e}_{t}\colon\hat{e}_{t+S-1}$ and $\hat{p}_{t+1}\colon\hat{p}_{t+S}$}
$\bm{c} \leftarrow\ \textrm{GRU}_\textrm{B}(f,\textrm{GRU}_\textrm{A}(f,\textrm{Embedding}(\textrm{Input})))$ \\
$\bm{c}_{t}^{h} = \bm{c}$ \\
\For{$i$ $\leftarrow$ $0$ \KwTo $S-1$}{
    $\hat{e}_{t+i}^{h} \leftarrow \textrm{sampling from}\ p(e_{t+i}^{h} | \bm{c}_{t+i}^{h})$ \\
    $\bm{c}_{t+i}^{l} \leftarrow \bm{c}_{t+i}^{h} + \textrm{Embedding}(\hat{e}_{t+i}^{h})$ \\
    $\hat{e}_{t+i}^{l} \leftarrow \textrm{sampling from}\ p(e_{t+i}^{l} | \bm{c}_{t+i}^{l})$ \\
    $\hat{e}_{t+i} \leftarrow 2^{B_{l}}\hat{e}_{t+i}^{h}+\hat{e}_{t+i}^{l}$ \\
    $\bm{c}_{t+i+1}^{h} \leftarrow \bm{c}_{t+i}^{h} + \textrm{Embedding}(\hat{e}_{t+i})$ \\
}
\For{$i$ $\leftarrow$ $0$ \KwTo $S-1$}{
    $\hat{s}_{t+i} \leftarrow \hat{e}_{t+i} + \hat{p}_{t+i}$ \\
    $\hat{p}_{t+i+1} \leftarrow \textrm{compute prediction}(\hat{s}_{t+i-M-1}\colon\hat{s}_{t+i})$ \\
}
\caption{Audio sample generation with sample and bit bunching}
\label{alg:bunch_sample_generation}
\end{algorithm}

As in the LPCNet paper, we have employed $\mu$-law quantization algorithm~\cite{mulaw-quantization} to represent 16-bits PCM values $x$, where $-32768 \leq x \leq 32767$. It is an efficient method which can represent the waveform with $B$-bits as shown in (\ref{original_mulaw}).
%
\begin{equation}
\begin{aligned}
y=Q_{B}(x) =& sign(x) \cdot V_{m2} \cdot \frac{\ln(1+s_{1}|x|)}{\ln(V_{m})} \\
x=Q_{B}^{-1}(y) =& sign(u) \cdot s_{2} \cdot \left(\exp\frac{\ln(V_{m})|u|}{V_{m2}}-1\right) \\
\text{where } V_{m}=&2^{B}, V_{m2}=2^{B-1}, \\
u=&y-V_{m2}, s_{1}=\frac{V_{m}-1}{2^{15}}, s_{2}=\frac{2^{15}}{V_{m}-1}
\end{aligned}
\label{original_mulaw}
\end{equation}

\begin{equation}
V_{m}=w_{s}2^{B}
\label{modified_mulaw}
\end{equation}

\begin{figure}[h]
\centering
\includegraphics[width=0.8\linewidth]{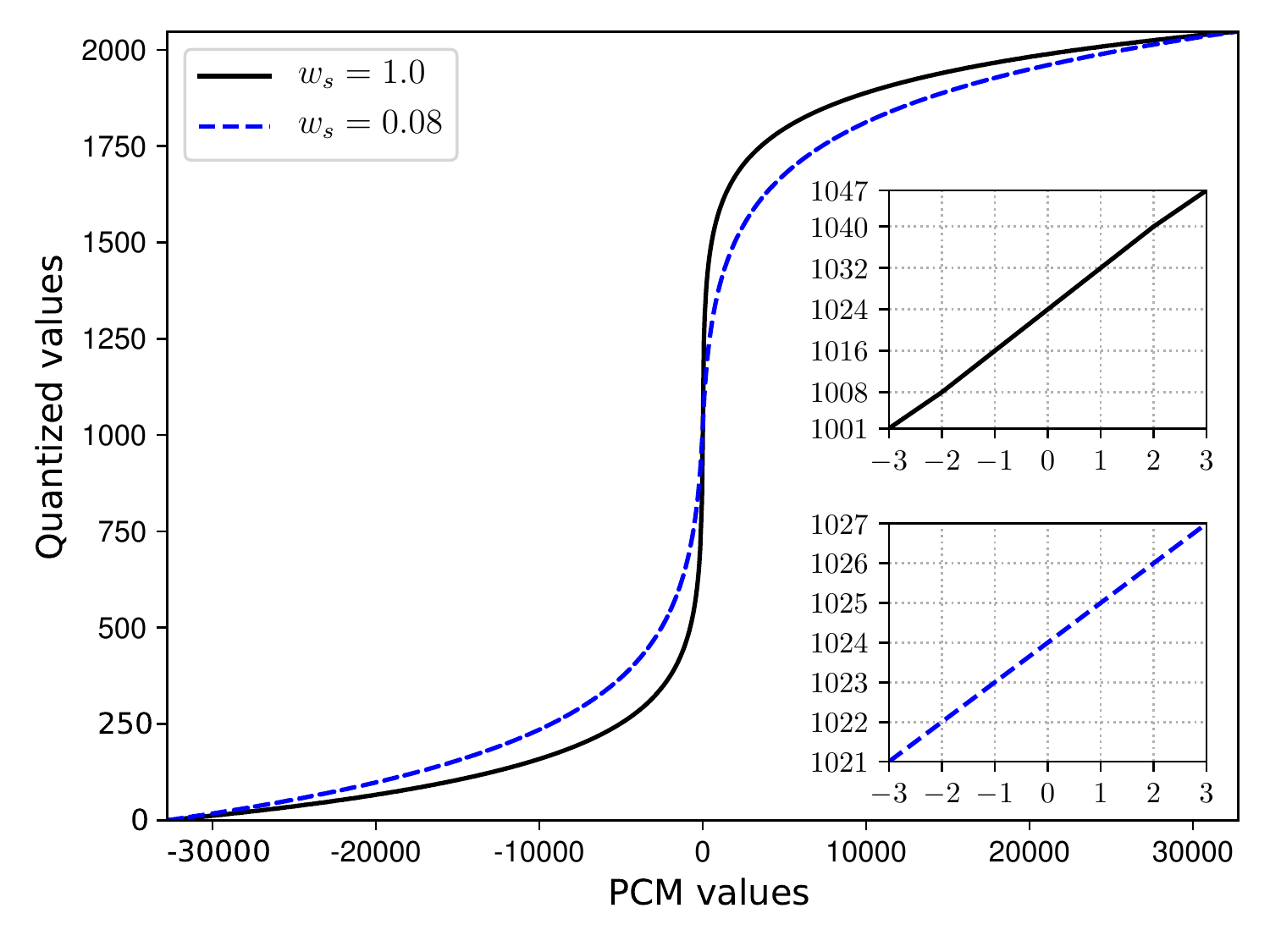}
\caption{Mapping function of the modified $\mu$-law quantization with $B=11$}
\label{fig:modified_mulaw_plot}
\end{figure}

Typically, most systems use it with $B=8$, but with bit bunching we consider cases with larger $B$. One downside with $B>9$ is that the quantization step size is smaller than 1 for values of $x$ close to zero. For instance for $B=11$, $Q_{11}(0)=1024$ and $Q_{11}(1)=1032$. This leads to under-utilization of quantization levels and also discrepancies due to many-to-one mapping from sampled quantization value to the PCM conversion. To address this issue, we added a factor $w_{s}$ to control the slope of the mapping function as represented in (\ref{modified_mulaw}), and it was chosen to keep the quantization step always greater than 1. For instance, for $B=11$, $w_{s}$ is set to 0.08 as shown in Figure \ref{fig:modified_mulaw_plot}.

\section{Evaluation}




\subsection{Experimental environment}
For high-fidelity TTS with low complexity, we made the following modifications to the original LPCNet configuration: (1) For 24KHz sampling rate, 10 ms frame size ($N$ : 240), 20 Bark cepstral coefficients with 240 shift size and 480 window size are used. (2) We used the RAPT algorithm \cite{rapt} for pitch tracking. It showed the better performance in the LPCNet and in generation of more natural prosody in the acoustic model. (3) We increased the sparse ratio of the recurrent weight matrix in $\text{GRU}_{\text{A}}$ from (0.95, 0.95, 0.8) to (0.99, 0.99, 0.9). This way, the overall complexity was reduced by about 25\% with a similar performance, and we employed this modified LPCNet as a baseline system for all comparisons. The learning rate and decay rate were determined by random search method \cite{bergstra2012random} and the other hyper-parameters were set identical to the original LPCNet, i.e., $\text{GRU}_{\text{A}}$ with 384 units and $\text{GRU}_{\text{
B}}$ with 16 units. A total of 8 systems were evaluated with the sample bunching sizes $S \in \{1,2,3,4\}$ and the bit bunching configurations $\bm{B}=(B_{h},B_{l}) \in \{(8,0), (7,4) \}$. Note that the system with $S=1$ and $\bm{B}=(8,0)$ corresponds to the baseline system. When $\bm{B}=(7,4)$, the modified $\mu$-law quantization was applied with $w_{s}=0.08$.

A phoneme-based DCTTS~\cite{tachibana2018efficiently} was employed as an acoustic model to evaluate TTS performance. The systems were trained using two datasets, a professional English male speaker (17-hours with 10,000 utterances) and a professional English female speaker (15-hours with 7,612 utterances). 110 utterances were used as a test set, and one percent of the rest were held out as validation set for training. 


\subsection{Complexity}
We measured RTF (Real Time Factor) and CR (Complexity Ratio) on two devices: 1) AWS c5.4xlarge instance - representative of cloud deployment (Intel(R) Xeon(R) Platinum 8124M CPU @ 3.00GHz) and 2) Samsung Galaxy S10+ (Exynos 9820) for on-device deployment.
The implementation was optimized using SIMD (Single Instruction Multiple Data) with single thread for each architecture.


\begin{table}[th]
	\caption{Real time factor and complexity ratio on the two CPU architectures}
	\label{tab:RTF}
	\centering
	\begin{tabular}{ cccccc }
		\toprule
		\multirow{2}{*}{$\bm{S}$} &
		\multirow{2}{*}{$\bm{B}$} &
		\multicolumn{2}{c}{\textbf{Intel Xeon}} &
		\multicolumn{2}{c}{\textbf{Exynos 9820}} \\
		& & \textbf{RTF} & \textbf{CR} & \textbf{RTF} & \textbf{CR} \\
		\midrule
		\multirow{2}{*}{$1$} & $(8,0)$ & $0.136$ & $100.0 \%$ &  $0.243$ & $100.0 \%$ \\
		 & $(7,4)$ & $0.127$ & $93.4 \%$ &  $0.214$ & $88.3 \%$ \\
		\midrule
		\multirow{2}{*}{$2$} & $(8,0)$ & $0.098$ & $72.1 \%$ &  $0.174$ & $71.7 \%$ \\
		 & $(7,4)$ & $0.089$ & $65.4 \%$ &  $0.149$ & $61.2 \%$ \\				\midrule
		\multirow{2}{*}{$3$} & $(8,0)$ & $0.087$ & $64.0 \%$ &  $0.147$ & $60.6 \%$ \\
		 & $(7,4)$ & $0.078$ & $57.4 \%$ &  $0.124$ & $51.2 \%$ \\				\midrule
		\multirow{2}{*}{$4$} & $(8,0)$ & $0.082$ & $60.3 \%$ &  $0.137$ & $56.6 \%$ \\
		 & $(7,4)$ & $0.072$ & $52.9 \%$ &  $0.111$ & $45.8 \%$ \\
		\bottomrule
	\end{tabular}
\end{table}

The RTFs and CRs summarized in Table~\ref{tab:RTF} highlight the improvements in computational complexity with sample and bit bunching over the baseline : $\bm{S}=1$ and $\bm{B}=(8, 0)$. The sample bunching algorithm with $S=4$ achieves a reduction of $39.7\%$ and $43.4\%$ over the baseline on Intel Xeon and Exynos respectively. Bit bunching helps in reducing the RTF further by about $7\%$ and $10\%$ absolute on these devices. With $\bm{S}=4$ and $\bm{B}=(7, 4)$, we get an overall improvement of $54.2\%$, i.e., it runs 2.19 times faster than the baseline, on the mobile device.


Since we need one Dual FC layer computation per excitation calculation even with sample bunching, it implies a constant reduction in computations with bit bunching for all values of $S$. At higher sample bunching values, this accounts for a larger proportion of savings, for e.g, relative gain with bit bunching on Intel Xeon at baseline is 6.6\% while at $S=4$, it results in 12.2\% gain.

\begin{figure}[h]
	\centering
	\includegraphics[width=0.85\linewidth]{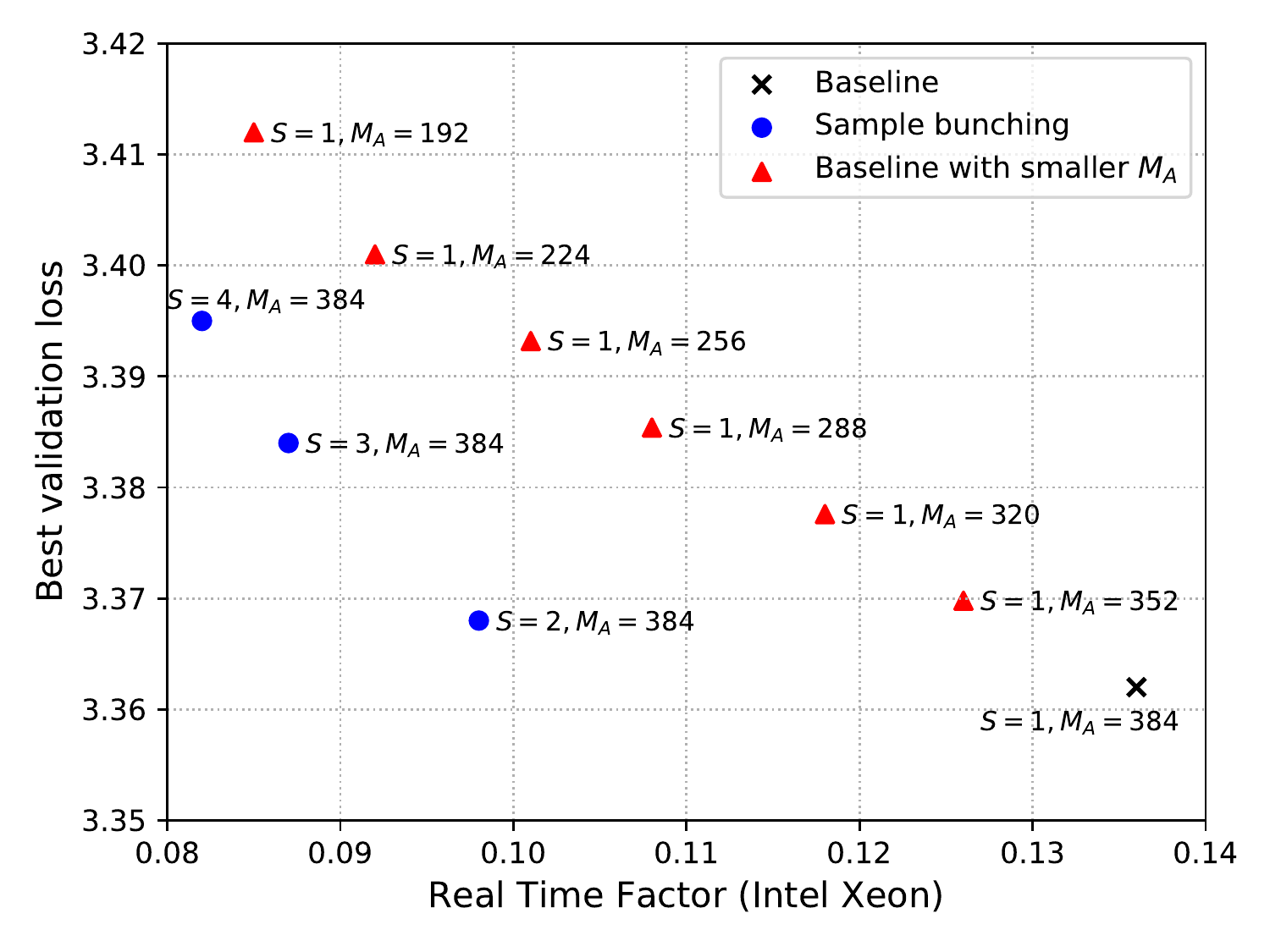}
	\caption{Validation loss versus RTF (baseline systems with varying $M_{A}$ and sample bunching systems)}
	\label{fig:comp_loss}
\end{figure}

To verify the efficacy of the bunching approach in complexity reduction, we compare the validation loss of the proposed system and the baseline system with smaller $\text{GRU}_{\text{A}}$ units $M_{A}$ in Figure~\ref{fig:comp_loss}. Sample bunching shows lower complexity for the same validation loss. For example, $(S=1, M_{A}=288)$ and $(S=3, M_{A}=384)$ with similar loss, 3.3854 and 3.3840, work 25.9\% and 56.3\% faster than the baseline system $(S=1, M_{A}=384)$ respectively. It suggests that the proposed approach is a more efficient method than reducing the units in the RNN layers.

\subsection{Quality}
For quality evaluation, we conducted a MOS (Mean Opinion Score) test and a DMOS (differential MOS) test using the Amazon Mechanical Turk platform on 11 systems including one original speech as a high anchor, and two degraded ones generated by Griffin-Lim \cite{griffin1984signal} and 5 bits $\mu$-law quantization as low anchors, with 100 people and 100 unseen test utterances. Each listener rated 110 utterances, 10 utterances each for the 11 systems.

For the evaluation, we choose Degradation Category Rating (DCR) manners for \textit{Anal-Synth} and Absolute Category Rating (ACR) manners for \textit{LPCNet+DCTTS} test defined in ITU-T P.800~\cite{itu-t}, for precise analysis of the vocoder performance. In the test, the subjects are requested to rate the amount of degradation compared with given reference in case of \textit{Anal-Synth} and to rate absolute quality in the case of \textit{LPCNet+DCTTS}. There are 10 categories in our test [0.5, 5] with a step size of 0.5 and a higher score corresponds to better quality. To remove non-discriminative ratings, we applied a screening rule where results were discarded if the original item scored lower than 4.0. The results of \textit{Anal-Synth} using the extracted features from original speech and \textit{LPCNet+DCTTS} using the predicted features from the DCTTS are summarized in Tables~\ref{tab:DMOS_analsynth} and \ref{tab:MOS_dctts} respectively\footnote{Audio samples available at \url{https://bunchedlpcnet.github.io/}}.

\begin{table}[th]
	\caption{DMOS scores with 95\% confidence intervals}
	\label{tab:DMOS_analsynth}
	\centering
	\begin{tabular}{ cccccc }
		\toprule
		\multirow{2}{*}{$\bm{S}$} &
		\multirow{2}{*}{$\bm{B}$} &
		\multicolumn{2}{c}{\textbf{\textit{Anal-Synth} (DMOS)}} \\
		& & \textbf{male} & \textbf{female} \\
		\midrule
		\multicolumn{2}{c}{\textbf{Original}} & $4.62 \pm0.03$ & $4.66 \pm0.03$ \\
		\midrule
		\multirow{2}{*}{$1$} & $(8,0)$ & $4.35 \pm0.06$ & $4.40 \pm0.06$ \\
		& $(7,4)$ & $4.32 \pm0.06$ & $4.39 \pm0.06$ \\
		\midrule
		\multirow{2}{*}{$2$} & $(8,0)$ & $4.30 \pm0.06$ & $4.30 \pm0.06$\\
		& $(7,4)$ & $4.25 \pm0.06$ & $4.29 \pm0.06$\\
		\midrule
		\multirow{2}{*}{$3$} & $(8,0)$ & $4.26 \pm0.06$ & $4.25 \pm0.07$ \\
		& $(7,4)$ & $4.18 \pm0.06$ & $4.20 \pm0.06$\\
		\midrule
		\multirow{2}{*}{$4$} & $(8,0)$ & $4.22 \pm0.06$ & $4.09 \pm0.07$\\
		& $(7,4)$ & $4.16 \pm0.06$ & $4.00 \pm0.07$\\
		\bottomrule
	\end{tabular}
\end{table}

\vspace*{-\baselineskip}
\begin{table}[th]
	\caption{MOS scores with 95\% confidence intervals}
	\label{tab:MOS_dctts}
	\centering
	\begin{tabular}{ cccccc }
		\toprule
		\multirow{2}{*}{$\bm{S}$} &
		\multirow{2}{*}{$\bm{B}$} &
		\multicolumn{2}{c}{\textbf{\textit{LPCNet+DCTTS} (MOS)}} \\
		& & \textbf{male} & \textbf{female} \\
		\midrule
		\multicolumn{2}{c}{\textbf{Original}} &  $4.41 \pm0.03$ & $4.44 \pm0.03$ \\
		\midrule
		\multirow{2}{*}{$1$} & $(8,0)$ &  $4.08 \pm0.05$ & $4.09 \pm0.05$ \\
		& $(7,4)$ &  $4.02 \pm0.05$ & $4.10 \pm0.05$ \\
		\midrule
		\multirow{2}{*}{$2$} & $(8,0)$ &  $4.05 \pm0.05$ & $4.06 \pm0.05$ \\
		& $(7,4)$ &  $4.04 \pm0.05$ & $4.06 \pm0.05$ \\
		\midrule
		\multirow{2}{*}{$3$} & $(8,0)$ &  $4.01 \pm0.06$ & $4.09 \pm0.05$ \\
		& $(7,4)$ &  $4.00 \pm0.05$ & $4.06 \pm0.05$ \\
		\midrule
		\multirow{2}{*}{$4$} & $(8,0)$ &  $4.01 \pm0.06$ & $4.06 \pm0.05$ \\
		& $(7,4)$ &  $3.99 \pm0.05$ & $4.03 \pm0.06$ \\
		\bottomrule
	\end{tabular}
\end{table}





In the \textit{Anal-Synth} case, it is confirmed that lower the complexity, lower is the DMOS. Especially, DMOS decreases rapidly when $S=4$, which can be attributed to the insufficient capacity of $\text{GRU}_{\text{B}}$. On the other hand, in the TTS case, MOS degradation is insignificant, and the proposed method with $S=4$ and $B=(7,4)$, shows a less than 0.1 MOS drop compared to the baseline. In fact, DCR test is more sensitive than ACR and the difference according to the test methodology shows up in the DMOS and MOS score of original utterance. 

\section{Conclusion}
In this work, we have proposed sample and bit bunching techniques to reduce the computational complexity of the LPCNet vocoder. We have demonstrated that bunched-LPCNet can provide more than 2X speed-up over baseline with negligible loss in quality. It is thus a strong proposition for use within a commercial on-device TTS system for lower-spec mobile, IoT and embedded devices.




\bibliographystyle{IEEEtran}

\bibliography{mybib}

\end{document}